\documentclass[aps,twocolumn,preprintnumbers,groupedaddress,superscriptaddress,floatfix,
tightenlines,reprint,longbibliography,nofootinbib]{revtex4-1}
\usepackage{mathrsfs}
\usepackage{natbib}
\usepackage{verbatim}
\usepackage{enumerate}
\usepackage{graphicx,bbm,amsbsy,amsfonts,amssymb,amsmath}
\usepackage{bm}
\usepackage{hyperref}
\usepackage{booktabs}
\usepackage{gensymb}

\usepackage{slashed}

\usepackage{footmisc}

\usepackage{blindtext}            
 
\hypersetup{
    colorlinks=true,       
    linkcolor=red,          
    citecolor=blue,        
    filecolor=magenta,      
    urlcolor=blue           
}

\def\ZZ{\mathbb{Z}} 
\def\RR{\mathbb{R}}

\def\2{{(2)}}
\def\1{{(1)}}
\def\0{{(0)}}
\def\m1{{(-1)}}

\newcommand{\transpose}{\ensuremath{{}^{\top}}}

\let\oldAA\AA
\renewcommand{\AA}{\text{\normalfont\oldAA}}

\newcommand{\Piphi}{\mathbf{\Pi_{\varphi}}}
\newcommand{\Pia}{\mathbf{\Pi}_a}

\newcommand{\phiop}{\boldsymbol{\varphi}}
\newcommand{\chiop}{\boldsymbol{\chi}}
\newcommand{\nop}{\mathbf{n}}
\newcommand{\qop}{\mathbf{q}}
\newcommand{\Qop}{\mathbf{Q}}
\newcommand{\Hop}{\mathbf{H}}
\newcommand{\aop}{\mathbf{a}}
\newcommand{\Gphi}{\mathbf{G}_{\varphi}}
\newcommand{\Gchi}{\mathbf{G}_{\chi}}
\newcommand{\Gsmall}{\mathbf{G}_{\rm small}}
\newcommand{\Glarge}{\mathbf{G}_{\rm large}}
\newcommand{\Uchiral}{\mathbf{U}}
\newcommand{\Voneform}{\mathbf{V}}

\newcommand{\ts}{\tilde{s}}
\newcommand{\tell}{\tilde{\ell}}

\begin{document}

\title{Exact lattice chiral symmetry in 2d gauge theory}
\author{Evan Berkowitz}
\email{e.berkowitz@fz-juelich.de }
\affiliation{Institut f\"{u}r Kernphysik \& Institute for Advanced Simulation,
Forschungszentrum J\"{u}lich, 54245 J\"{u}lich, Germany
}
\affiliation{Center for Advanced Simulation and Analytics (CASA),
Forschungszentrum J\"{u}lich, 52425 J\"{u}lich, Germany
}
\author{Aleksey Cherman}
\email{acherman@umn.edu}
\affiliation{School of Physics and Astronomy, University of Minnesota, Minneapolis, MN 55455, USA}
\author{Theodore Jacobson}
\email{tjacobson@physics.ucla.edu}
\affiliation{Mani L. Bhaumik Institute for Theoretical Physics, Department of Physics and Astronomy, University of California, Los Angeles, CA 90095, USA}

\begin{abstract}
We construct symmetry-preserving lattice regularizations of 2d QED with one and
two flavors of Dirac fermions, as well as the ``$3450$'' chiral gauge theory, by
leveraging bosonization and recently-proposed modifications of Villain-type
lattice actions. The internal global symmetries act just as locally on the
lattice as they do in the continuum, the anomalies are reproduced at finite
lattice spacing, and in each case we find a sign-problem-free dual formulation.
\end{abstract}

\maketitle
\flushbottom

\section{Introduction}  
Numerical Monte Carlo simulations of quantum field theories (QFTs) discretized
on Euclidean spacetime lattices are one of the few known non-perturbative
techniques to study strongly coupled QFTs. However, it is famously difficult to
discretize fermions while preserving all of their
symmetries~\cite{Kaplan:2009yg}. For example, a free massless Dirac fermion has
the internal global symmetry $\left[(U(1)_V \times U(1)_A)/\ZZ_2 \right] \rtimes
(\ZZ_2)_C$ for even $d$.  The continuous symmetries have a mixed 't Hooft
anomaly, and standard lattice regularizations do not preserve the continuum
version of the chiral symmetry at finite lattice spacing $a$.

If we integrate out a massless Dirac fermion in a Euclidean QFT, we obtain the
path integral
\begin{align}
  Z = \int \mathcal D\phi \det\left[ \slashed{D}(\phi)\right] e^{-S(\phi)}
\end{align}
where $\slashed{D}(\phi) = \gamma^{\mu} D_{\mu}(\phi)$ is the Dirac operator and
$\phi$ stands for an appropriate set of bosonic fields with path integral
measure $\mathcal D\phi$ and Euclidean action $S(\phi)$. The starting point for
lattice Monte Carlo studies is a discretization of $Z$ that preserves as much of
the internal symmetry of the QFT as possible.

Replacing the massless continuum Dirac operator by a simple lattice difference
operator on a (hyper)-cubic lattice does not give the desired symmetries and
anomalies. Instead, it yields $2^d$ massless Dirac
fermions in the continuum limit with the
symmetry charges of the `doubler' fermions such that the chiral anomaly
cancels~\cite{Karsten:1980wd,Kawamoto:1981hw}. The Nielsen-Ninomiya
theorem~\cite{Nielsen:1980rz,Nielsen:1981hk,Nielsen:1981xu,Friedan:1982nk}
states that in fact there is no lattice Dirac operator which is simultaneously
local, has the desired continuum limit with just one massless Dirac fermion, and
is consistent with a locally-acting chiral symmetry $\{\Gamma, \slashed{D}\} =
0$ where $\{\Gamma, \gamma^{\mu}\} = 0$.

The standard ways around this `fermion doubling problem' all give up some
desirable features of the continuum theory. Wilson fermions remove the doublers
but explicitly break chiral
symmetry~\cite{Wilson1977,Karsten:1980wd,Kawamoto:1981hw}. Staggered
fermions~\cite{Kogut:1974ag,Susskind:1976jm,Banks:1975gq,SHARATCHANDRA1981205,Burden:1986by}
do not remove all the doublers.\footnote{However, when the continuum theory of
interest has the same number of fermions as produced via doubling, one can use
staggered fermions (or the closely related Kahler-Dirac fermions) to reproduce
the anomalies of the continuum theory, see e.g.
\cite{Catterall:2022jky,Catterall:2023nww}.} Domain-wall and overlap
fermions~\cite{Kaplan:1992bt,Shamir:1993zy,Furman:1994ky,Narayanan:1993ss,
Narayanan:1993sk,Narayanan:1994gw,Neuberger:1997fp,Neuberger:1998wv,Clancy:2023ino},
which satisfy\footnote{In the case of domain-wall fermions the Ginsparg-Wilson
relation is satisfied in the limit where the extra dimension is infinitely
large.} the Ginsparg-Wilson relation $\{\Gamma, \slashed{D} \} = a
\slashed{D}\Gamma \slashed{D}$~\cite{Ginsparg:1981bj}, remove all of the
undesired doubler modes at the cost of making both chiral symmetry
transformations and the  Dirac operator itself non-local at finite lattice
spacing~\cite{Luscher:1998pqa}.

This was historically viewed as an unavoidable consequence of anomalies, which
in popular textbook presentations are characterized as solely arising from
subtleties in regularizing fermions. Relatedly, there is a belief that 't Hooft
anomalies are necessarily absent in lattice theories with locally acting
symmetries~\cite{Nielsen:1980rz,Karsten:1980wd,Nielsen:1981hk,Ginsparg:1981bj,Kaplan:2009yg},
so that the overlap formulation is the best one can
do~\cite{Kaplan:1992bt,Shamir:1993zy,Furman:1994ky,Narayanan:1993ss,
Narayanan:1993sk,Narayanan:1994gw,Neuberger:1997fp,Neuberger:1998wv,Clancy:2023ino}.

However, anomalies are not restricted to fermionic systems, and it has recently
become appreciated that there exist lattice discretizations in which anomalies
of locally-acting symmetries can appear even at finite lattice
spacing~\cite{Sulejmanpasic:2019ytl,
Gorantla:2021svj,Fazza:2022fss,Nguyen:2022aaq,Cheng:2022sgb,Seifnashri:2023dpa}.
We show that these results straightforwardly lead to lattice discretizations of
Dirac fermions coupled to abelian gauge fields in $d=2$ which preserve the
internal symmetries and anomalies \emph{exactly}, with chiral symmetries acting
locally even at finite lattice spacing. Our approach is to first apply abelian
bosonization to $N_f$ Dirac fermions, and then discretize the resulting bosonic
theory using an appropriate modified Villain action.\footnote{A more
conventional discretization of the bosonized Schwinger model was studied in
Refs.~\cite{Ohata:2023sqc,Ohata:2023gru}. Here our main focus is on the
symmetries and global aspects of the model, and our analysis leverages a number
of special features of the modified Villain formalism. An alternative approach
to discretization of  bosonized 2d gauge theories that shares some (but not all)
features of the modified Villain construction was recently discussed in
Ref.~\cite{DeMarco:2023hoh}. } We discuss how this works in 2d QED with $N_f= 1$
and $N_f = 2$ charge $Q$ fermions and in the ```$3450$''' abelian chiral gauge
theory. We also discuss a related spatial lattice Hamiltonian for $N_f=1$
QED in Appendix~\ref{sec:HamiltonianAppendix}.

{\bf Bosonization.} Consider the charge $Q$ Schwinger model: 2d QED with a
massless Dirac fermion coupled to a $U(1)$ gauge field $a_{\mu}$ with electric
charge $Q\in\ZZ$~\cite{Schwinger:1962tn,Schwinger:1962tp,Coleman:1975pw,
Coleman:1976uz,Iso:1988zi,Gross:1995bp,Anber:2018jdf,Anber:2018xek,
Misumi:2019dwq,Funcke:2019zna,Komargodski:2020mxz,Cherman:2020cvw,Cherman:2021nox,
Cherman:2022ecu,Honda:2022edn}. We normalize $a_{\mu}$ such that
$\frac{1}{4\pi}\int_M d^2 x\, \epsilon^{\mu\nu} f_{\mu \nu} \in \ZZ$, where
$f_{\mu\nu} = \partial_{\mu} a_{\nu} - \partial_{\nu} a_{\mu}$, and write the
action as
\begin{align}
    S
    =
    \int d^2x  \left[
        \frac{1}{4e^2} f_{\mu\nu}f^{\mu\nu} 
    +   \bar{\psi} \gamma^{\mu}(\partial_{\mu} - i Q a_{\mu}) \psi
    \right].
    \label{eq:one_flavor_QED}
\end{align}
The Nielsen-Ninomiya theorem constrains discretizations of $\slashed{D}$, but
does not directly constrain $\det \slashed{D}$. We thus aim to circumvent this
theorem by discretizing $\det \slashed{D}$ directly, by using the fact that in
$d=2$~\cite{Coleman:1974bu,Coleman:1975pw,PhysRevD.11.2088,Coleman:1976uz}
\begin{align} \label{eq:bosonized_action}
  \det(\slashed{D}(a_{\mu})) &=\\
   \int \mathcal D \varphi \exp\bigg[&-\int d^2x\, 
  \bigg(\frac{1}{8\pi}\partial_{\mu}\varphi \partial^{\mu}\varphi  
  + \frac{i Q}{2\pi} \epsilon^{\mu\nu} a_{\mu}\partial_{\nu} \varphi \bigg)\bigg] \,.
 \nonumber
\end{align}
In this `bosonized' action $\varphi$ is a compact real scalar field
$\varphi\equiv \varphi +2\pi$ and the mapping of the $U(1)_V, U(1)_A$ currents
is $\bar{\psi} \gamma^{\mu} \psi \leftrightarrow  - \frac{1}{2\pi}
\epsilon^{\mu\nu}\partial_{\nu} \varphi$, $ \bar{\psi} \gamma^{\mu} \gamma^5
\psi \leftrightarrow   \frac{i}{4\pi}\partial_{\mu} \varphi$. We hasten to
emphasize that the existence of a map to bosonic variables
\eqref{eq:bosonized_action} does \emph{not} mean that the fermion discretization
problem is trivially solvable. Such a solution requires exhibiting a lattice
action in which all the desired symmetries and anomalies are preserved.

The ABJ anomaly is encoded at tree level in \eqref{eq:bosonized_action}, where
it is clear that the $0$-form symmetry counting chiral charges of local
operators is $(\ZZ_Q)_A$,  acting as $\varphi \to \varphi +2\pi k/Q$, rather
than $U(1)_A$. There is also a $1$-form~\cite{Gaiotto:2014kfa} `electric'
symmetry $(\ZZ_Q)_{\rm e}$ which counts the charges of Wilson loops modulo $Q$,
as well as a mixed 't Hooft anomaly between $(\ZZ_{Q})_{A}$ and $(\ZZ_Q)_{\rm
e}$ which is matched by the spontaneous breaking of \emph{both} symmetries, with
the walls separating chiral vacua carrying electric
charge~\cite{Anber:2018jdf,Anber:2018xek,Misumi:2019dwq,Komargodski:2020mxz,
Cherman:2020cvw,Cherman:2021nox,Cherman:2022ecu}. The spectrum in each
degenerate discrete chiral vacuum consists of a single free massive scalar field
with mass $m_\gamma = eQ/\pi$, often called the Schwinger boson.

\section{Modified Villain discretization}

\begin{figure}[h]
  \centering
  \includegraphics[width=0.3\textwidth]{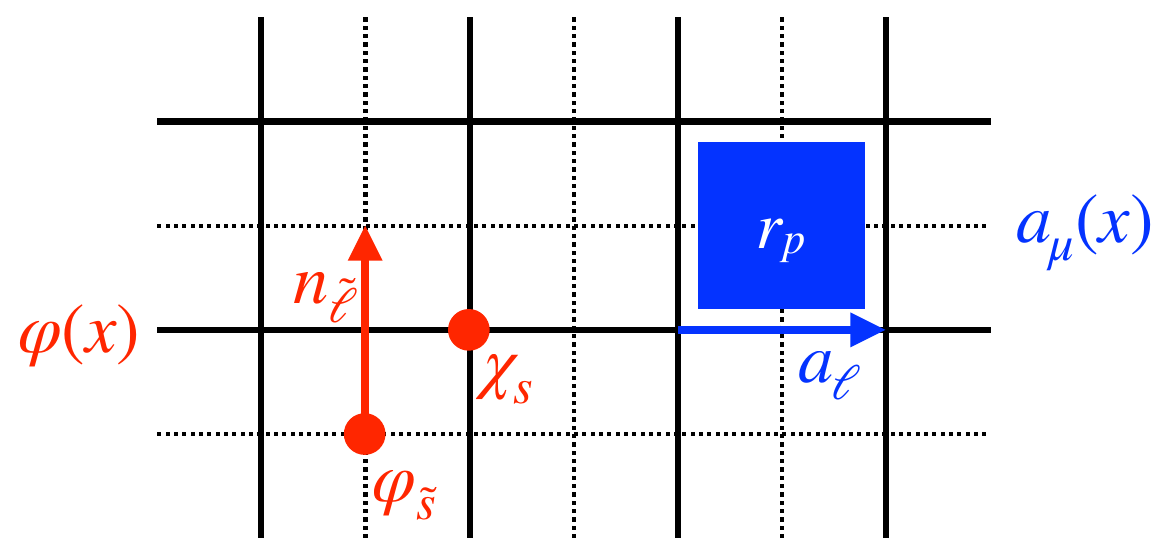}
  \caption{ The setting for the field content of our lattice action
    \eqref{eq:lattice_action}. The solid grid is the primary lattice with sites
    $s$, links $\ell$, and plaquettes $p$. The dotted grid is the dual lattice
    with sites $\tilde{s}$, links $\tilde{\ell}$, and plaquettes $\tilde{p}$.
    The three red fields $\varphi$, $\chi$, and $n$ are associated with the
    continuum $\varphi$. The two blue fields $a$ and $r$ correspond to the
    continuum $a_\mu$. }
  \label{fig:villain_discretization_figure}
\end{figure}

We will work with an $N \times N$ periodic Euclidean spacetime lattice with
spacing $a=1$, with sites $s$, links $\ell$, and plaquettes $p$. The
corresponding simplices on the dual lattice are denoted by $\tilde{s},
\tilde{\ell}, \tilde{p}$. Following Villain~\cite{Villain:1974ir}, we represent
the continuum $U(1)$ gauge field $a_{\mu}$ by a pair of lattice fields $\{
a_{\ell} \in \RR, r_{p} \in \ZZ\}$ and the compact scalar field by the pair
$\{\varphi_{\tilde s} \in \RR, n_{\tilde{\ell}} \in \ZZ\}$ on the dual lattice.
We adopt the modified~\cite{Sulejmanpasic:2019ytl,Gorantla:2021svj} Villain
formulation, and also introduce an auxiliary field $\chi_{s} \in \RR$ which can
be viewed as the T-dual of $\varphi_{\tilde s}$. See
Fig.~\ref{fig:villain_discretization_figure} for an illustration.

The action for our discretization of $N_f=1$ QED is
\begin{align}\label{eq:lattice_action}
    S_{N_f = 1}
    &=
        \frac{\beta}{2} \left[(da)_{p} - 2\pi r_p\right]^2
    +   \frac{\kappa}{2} \left[(d\varphi)_{\tilde{\ell}} - 2\pi n_{\tilde{\ell}}\right]^2 
   \\
    &
    -   i \chi_s (dn)_{\star s} 
    +   \frac{i Q}{2\pi}\varphi_{\star p}  \left[(da)_p - 2\pi r_p \right] 
    -   i Q a_{\ell} \, n_{\star \ell}  
    \nonumber
\end{align}
where repeated indices are summed and $d$ is the lattice exterior derivative $(d
\omega)_{c^{r+1}} = \sum_{c^{r} \in \partial c^{r+1}} \omega_{c^r} $ where $c^r$
is an $r$-cell, so that, for example, $(d \chi)_{\ell} =\chi_{s+\hat{\ell}} -
\chi_s$, and $d^2 = 0$. The Hodge star $\star$ maps an $r$-cell $c^r$ on the
lattice to the $(d-r)$-cell $(\star c)^{d-r}$ on the dual lattice which pierces
$c^r$.\footnote{Differential forms on the lattice are reviewed in Appendix A of
Ref.~\cite{Sulejmanpasic:2019ytl}. Two useful facts are that $\star^2 =
(-1)^{r(d-r)}$ on an $r$-cell, and the identity $\sum_{c^{r+1}}
(dA)_{c^{r+1}}B_{\star c^{r+1}} = (-1)^{r+1} \sum_{c^r} A_{c^r} (dB)_{\star
c^r}\,.$ }  The partition function is
\begin{align}
Z= \prod_{s, \tilde{s}} \int_{\mathbb{R}} 
\mathcal{D}\chi_s \mathcal{D}\varphi_{\tilde{s}} 
\prod_{\ell} \int_{\mathbb{R}} \mathcal{D}a_{\ell} 
\sum_{n_{\ell} \in \mathbb{Z}} \prod_{p} \sum_{r_p \in \mathbb{Z}}
e^{-S_{N_f=1}}
\end{align}
where the products are over all sites, links, and dual sites of our periodic
square lattice with $N \times N$ sites.    Similar expressions can be written
for the other gauge theories considered in this paper.  The
$\mathbb{Z}$ gauge redundancy on each site makes $Z$ formally infinite, and
readers who find this uncomfortable can work in the `Villain' gauge where
integrals over $\mathbb{R}$ are replaced by integrals over the interval $(-\pi,
\pi]$, see e.g. Refs.~\cite{Sulejmanpasic:2019ytl,Anosova:2022cjm} for
discussions of various possible gauge choices.  However, $Z$ is in any case not
an observable, and all physical observables are necessarily finite even without
this gauge choice.

The gauge redundancies of the lattice action \eqref{eq:lattice_action} are
\begin{subequations}
\label{eq:one_flavor_gauge}
\begin{align}
    a_{\ell} &\to a_{\ell} + (d\lambda)_{\ell} + 2\pi m_{\ell}\,,
    \;
    r_p \to r_p + (dm)_p \\
    \varphi_{\tilde{s}} &\to \varphi_{\tilde{s}} + 2\pi k_{\tilde{s}}\,,
    \;
    n_{\tilde{\ell}} \to n_{\tilde{\ell}} + (dk)_{\tilde{\ell}}
    \\
    \chi_s &\to \chi_s + Q \lambda_s + 2\pi h_s
\end{align}
\end{subequations}
where $\{\lambda_s \in \RR, m_{\ell}, k_{\ts}, h_s \in \ZZ\}$ are gauge
parameters. They ensure that $\{a, r\}$ and $\{\chi,\varphi, n\}$ describe a
$U(1)$ gauge field and a $2\pi$-periodic boson with a conserved winding charge,
with the topological properties one expects in the continuum. For example, the
instanton number on the spacetime torus  $I = - \tfrac{1}{2\pi} \sum_{p}
\left[(da)_p - 2\pi r_p \right] = \sum_p r_p$ is an integer. The path integral
over $\chi_s$ implies that $(dn)_{\tilde{p}} = 0$ on shell, the $\frac{i
Q}{2\pi}$ term in the lattice action~\eqref{eq:lattice_action} is the analog of
the continuum $\frac{iQ}{2\pi}$ term~\eqref{eq:bosonized_action}, and the last
contribution to the lattice action \eqref{eq:lattice_action} is necessary to
maintain gauge invariance~\cite{Gorantla:2021svj}.

Given that $\beta=1/(2e^2 a^2)$, to get a continuum limit with fixed $Le$, where
$L$ is the physical box size $L = N a$, we should take $N \to \infty$ with
$\beta/N^2$ fixed.  While naively one should also set  $\kappa = 1/(4\pi)$ to
reach the continuum \eqref{eq:bosonized_action}, this parameter value is not
protected by any symmetries of the lattice theory, and can receive some finite
renormalization~\cite{Kosterlitz:1974sm,Villain:1974ir,Janke:1993va}. Varying
$\kappa$ amounts to varying the coefficient of the marginal Thirring term
$(\bar{\psi}\gamma^{\mu} \psi)^2$.

The lattice action \eqref{eq:lattice_action} has precisely the desired global
symmetries of the continuum \eqref{eq:one_flavor_QED}. There is no continuous
$U(1)_A$ symmetry, but thanks to the quantization of instanton number, there is
a remnant $(\ZZ_{Q})_{A}$ symmetry that acts as as $\varphi_{\tilde{s}} \to
\varphi_{\tilde{s}} +2\pi q/Q$ with $q \in \ZZ$. This reproduces the expected
ABJ chiral anomaly. The $(\ZZ_Q)_{\rm e}$ symmetry acts as $a_{\ell} \to
a_{\ell} + \frac{2\pi}{Q}v_\ell$ with $v \in \ZZ$ and $dv = 0$, also matching
the continuum.

To see the 't Hooft anomaly between $(\ZZ_Q)_{\rm e}$ and $(\ZZ_{Q})_{A}$ on the
lattice, it is easiest to linearize the quadratic terms in the action by
integrating in auxiliary fields $\zeta_{\ell}, \xi_{\tilde{s}} \in \RR$ and
summing by parts, turning the original action \eqref{eq:lattice_action}
into\footnote{Throughout the paper we ignore any overall constant factors in
the partition function which appear in `dualization' procedures.}
\begin{align}
  S'_{N_f =1}
    &=
    \left(
            \frac{1}{2\kappa} \zeta_{\ell}^2
        +   i \zeta_{\ell} \left[(d\varphi)_{\star\ell} - 2\pi n_{\star\ell}\right]
    \right)
    \label{eq:first_order_action}  
    \\
    &
    + 
    \left(
            \frac{1}{2\beta}  \xi_{\star p}^2
        +   i \xi_{\star p}  \left[(da)_{p} - 2\pi r_p\right]
    \right)
    \nonumber\\
    &
    + \frac{i Q}{2\pi} \varphi_{\star p} \left[(da)_p - 2\pi r_p \right] 
    - i n_{\star \ell}\left[Q a_{\ell} - (d\chi)_{\ell}\right] \,.
    \nonumber 
\end{align}
The generators of the axial $(\ZZ_Q)_{A}$ and electric $(\ZZ_Q)_{\rm e}$
symmetries are topological line and local operators on the lattice and dual
lattice, respectively:
\begin{equation}
  (U_A)[C] = e^{i \sum_{\ell \in C} \left( a_\ell + \frac{2\pi}{Q} \zeta_{\ell} \right)}\,, \;
  (U_{\rm e})_{\tilde s} = e^{i\left( \varphi_{\tilde s} + 
  \frac{2\pi}{Q}\xi_{\tilde s} \right)} \,,
\end{equation}
where $C$ is a closed curve. The fact that $(U_A)[C]$ is charged under
$(\ZZ_Q)_{\rm e}$ and $(U_{\rm e})_{p}$ is charged under $(\ZZ_Q)_{A}$ encodes
the mixed 't Hooft anomaly of these symmetries, just as in the continuum.

\section{Absence of a sign problem}
The discretizations provided above, and indeed, those presented below, provide
symmetry-preserving nonperturbative definitions of their respective models.
While this is interesting in its own right, it is natural to ask whether these
definitions have some practical value.  Can we learn something new about the
physics of these models just from defining them non-perturbatively, either
numerically or analytically?  On the analytic side, we will see in
Section~\ref{sec:3450_model} that putting a simple chiral gauge theory (the
`3450' model) on the lattice reveals the presence of an exotic symmetry of the
model that is not at all obvious from continuum analyses.   On the numerical
side, one might initially worry that the constructions described in this paper
are completely useless, because direct numerical Monte Carlo with the complex
lattice actions \eqref{eq:lattice_action} and \eqref{eq:first_order_action}
would face a severe sign problem. We now show that this issue can be eliminated
by a change of variables, so that the discretizations we give here can be
explored using numerical Monte Carlo simulations. 

Summing (i.e. path integrating) over $n_{\tilde \ell}, r_p$ in the auxiliary-field action
\eqref{eq:first_order_action} yields constraints that can be solved by setting
\begin{align}
  \zeta_{\ell} = \frac{1}{2\pi} (d\chi)_{\ell} - \frac{Q}{2\pi} a_{\ell} - u_{\ell}\,,\;
  \xi_{\tilde s} = - \frac{Q}{2\pi} \varphi_{\tilde{s}} + t_{\tilde{s}}
  \label{eq:constraint_sol_f_b}
\end{align}
with $u_{\ell}, t_{\tilde{s}} \in \ZZ$ which transform as $t_{\tilde{s}} \to
t_{\tilde{s}} + Q k_{\tilde{s}}$ and $u_{\ell} \to u_{\ell} + (dh)_{\ell} - Q
m_{\ell}$ under discrete gauge transformations. The field $u$ also transforms
under $(\ZZ_Q)_{\rm e}$ transformations $u_{\ell} \to u_{\ell} - 2\pi v_\ell/Q$,
while $t$ transforms under $(\ZZ_Q)_A$ as $t_{\tilde s} \to t_{\tilde s} +q$.
Plugging the constraints \eqref{eq:constraint_sol_f_b}  into the action
\eqref{eq:first_order_action} and dropping total derivatives and integer
multiples of $2\pi i$, we obtain an action
\begin{align}
    &   \frac{1}{2\kappa} \left[ \frac{1}{2\pi} (d\chi)_{\ell} - 
    \frac{Q}{2\pi} a_{\ell} - u_{\ell} \right]^2
    +   \frac{Q^2}{2\beta(2\pi)^2}  \left[\varphi_{\tilde{s}} -
    \frac{2\pi t_{\tilde{s}}}{Q} \right]^2 \nonumber \\
    &
    -   \frac{i}{2\pi} \left(Q a_{\ell} + 2\pi u_{\ell} \right) 
    (d\varphi)_{\star \ell} + i t_{\star p} (da)_p\,.
\end{align}
Shifting $a \to a + \frac{1}{Q} d\chi -\frac{2\pi}{Q} u$ and dropping a total
derivative gives
\begin{align}
 &\frac{1}{2\kappa} \left(\frac{Q}{2\pi} a_{\ell}\right)^2 
   + \frac{Q^2}{2\beta(2\pi)^2}  \left[\varphi_{\tilde{s}} - 
   \frac{2\pi t_{\tilde{s}}}{Q} \right]^2 \nonumber \\
   & - \frac{i}{2\pi} Q a_{\ell} (d\varphi)_{\star \ell} 
   + i t_{\star p} \left[(da)_p -\frac{2\pi}{Q} (du)_p\right]\,,
\end{align}
and subsequently integrating over $a$ yields
\begin{align}
  S_{N_f =1, \rm dual} &= \frac{\kappa}{2}  
  \left[(d\varphi)_{\tilde{\ell}} 
  - \frac{2\pi}{Q} (dt)_{\tilde{\ell}} \right]^2 \label{eq:real_dual_action} \\
  &+ \frac{1}{2\beta}\left(\frac{Q}{2\pi}\right)^2
  \left(\varphi_{\tilde{s}} - \frac{2\pi}{Q} t_{\tilde{s}}\right)^2
  - \frac{2\pi i}{Q} t_{\star p} (d u)_p
  \nonumber
\end{align}
This action describes $Q$ copies (`universes'
\cite{Hellerman:2006zs,Anber:2018jdf,Komargodski:2020mxz,Cherman:2020cvw,
Cherman:2021nox,Hellerman:2010fv,Sharpe:2014tca,Sharpe:2019ddn,
Robbins:2020msp,Sharpe:2021srf}) of a free massive scalar particle, as expected
from continuum arguments. Adding a fermion mass term in the original action
\eqref{eq:one_flavor_QED} corresponds to adding $\sum_{\tilde{s}}
\cos(\varphi_{\tilde{s}})$ to the dual action \eqref{eq:real_dual_action}, which
would lead to strong coupling in general. 

The sole imaginary term in the dual action \eqref{eq:real_dual_action} involves
$u$, but summing over $u$ just gives the constraint  $(dt)_{\tilde{\ell}} = 0
\textrm{ mod } Q$.  One can thus avoid the sign problem entirely by proposing
updates for $t_{\ts}$ that satisfy $(dt)_{\tilde{\ell}} = 0 \textrm{ mod } Q$ in a
Monte Carlo calculation.  

Procedures to generate field configurations which satisfy constraints  such as
$(dt)_{\tilde{\ell}} = 0 \textrm{ mod } Q$, and hence avoid apparent sign problems,
are well-known, see for example, Ref.~\cite{Gattringer:2018dlw}.  Nevertheless,
to make our presentation self-contained, we now give a brief discussion of how
simple constraints such as the one above --- and indeed others that we encounter
later in this paper --- can be taken into account in Monte Carlo calculations
without sign problems.  

Consider a lattice field theory with an action $S$ where the only term where the
field $u$ appears is
\begin{align}
    S \ni \frac{2\pi i }{Q} u_{\ell} (dt)_{\ell}
\label{eq:link_sign}
\end{align}
where $Q \in \mathbb{N}, u_{\ell} \in \mathbb{Z}, t_s \in \mathbb{Z}$. The dual
one-flavor \eqref{eq:real_dual_action}, two-flavor \eqref{eq:two_flavor_dual},
and 3450 \eqref{eq:dual_3450} models all have this character.

Because it is purely imaginary, direct Monte Carlo evaluation of the QFT path
integral based on importance-sampling with this term as part of the action
suffers from a sign problem. However, the path integral over $u$ can be done
analytically and yields a delta function setting 
\begin{align}
    (dt)_{\ell} = 0 \textrm{ mod } Q \,.
    \label{eq:simple_constraint}
\end{align} 
If we can make proposals that maintain this constraint but are otherwise ergodic
we will consider all supported configurations of $t_s$ and avoid the sign
problem caused by the phase \eqref{eq:link_sign}.

There is a simple solution to the constraint \eqref{eq:simple_constraint} on a
spacetime torus, where we can write\footnote{Formally, this very simple solution is
possible because all of the
integer cohomology groups of a torus are torsion-free.}
\begin{align}
    t_s = x + Q y_s
\end{align}
$x, y_s \in \mathbb{Z}$ and $x$ is a constant so that $(dx)_\ell = 0$.  This
decomposition is not unique, since (for example) we can shift $x$ by $Q$ and
all $y$'s by $-1$ without changing $t$.  But the key point is that any
constraint-satisfying $t_s$ can be written in the form above. This helps us
define two kinds of proposals which together reach all constraint-satisfying
configurations.

The first proposal is a global update of $x$. We randomly pick a
site-independent integer $\Delta x \in [-X, +X]$ with $X\in\ZZ$, and Metropolis
test $t_s \to t_s + \Delta x$ for all sites $s$ simultaneously.

The second is a local update of $y$ which we can sweep across the lattice. On a
particular site $s$ we pick an integer $\Delta y_s \in [-Y, +Y]$ with $Y\in\ZZ$,
and test $t_s \to t_s + Q \Delta y_s$.

An ergodic algorithm should offer proposals of both kinds, and their relative
frequency may be adjusted to control autocorrelation times. This algorithm also
manifestly satisfies detailed balance thanks to the Metropolis tests described
above. The algorithm parameters $X$ and $Y$, or more generally the distributions
for $\Delta x$ and $\Delta y_s$, may be adjusted to optimize acceptance and
thermalization.  

Similarly, suppose the action includes a term 
\begin{align}
    S \ni i \eta_{p} (dn)_{p}
\label{eq:link_plaquette}
\end{align}
where $\eta_p \in \mathbb{R}, n_{\ell} \in \mathbb{Z}$, and $\eta$ does not
appear in any other terms. The dual 3450 action \eqref{eq:dual_3450} has a
term of this character. Integrating out $\eta$ yields the constraint $(dn)_p=0$.

The field $n_\ell$ may be split into a closed $1$-form $w_\ell \in \ZZ$ and an
exact $1$-form $(dz)_\ell \in \ZZ$
\begin{align}
    n_\ell = w_\ell + (dz)_\ell.
\end{align}
As we will see below, updates of $z$ are local, while updates of $w$ `wrap'
around cycles of the torus. Again, for any given field configuration $n_{\ell}$, the
decomposition above is not unique, but the key point is that any
constraint-satisfying $n_{\ell}$ can be written in the form above.

\begin{figure}[h]
    \centering
    \includegraphics[width=0.3\textwidth]{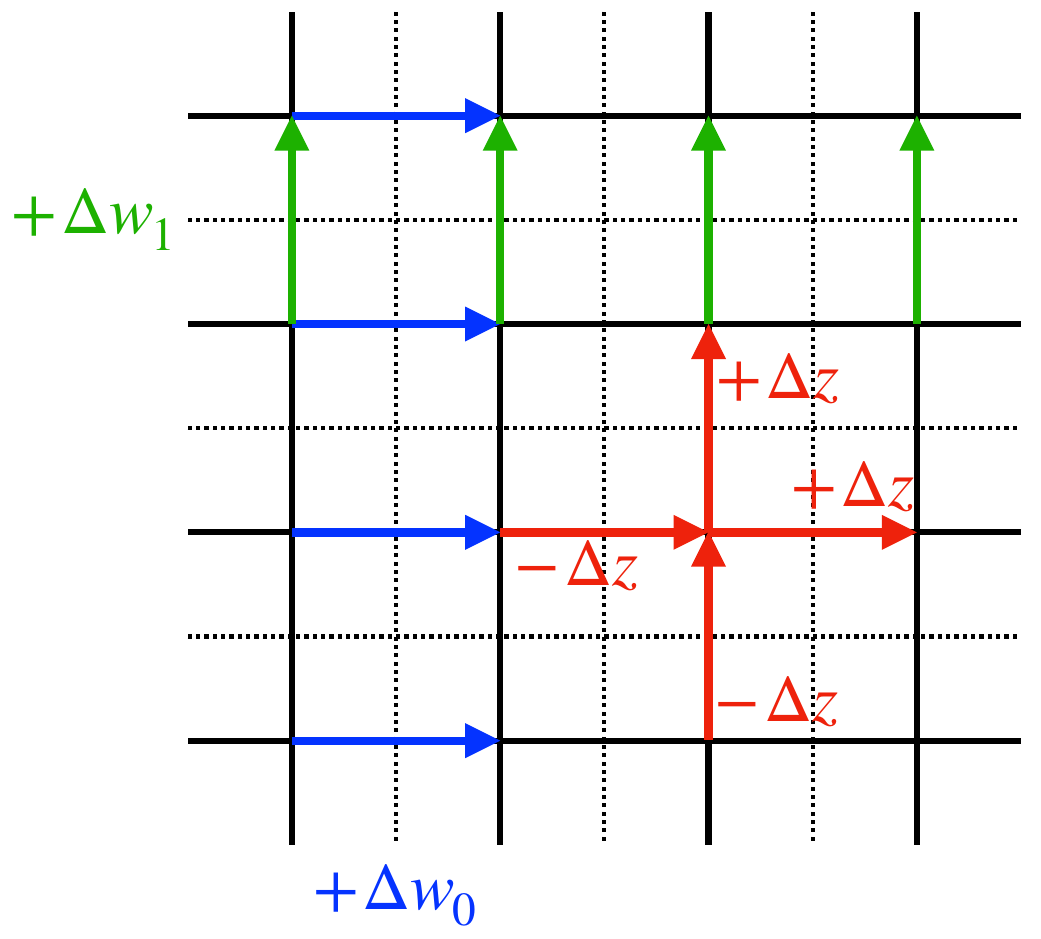}
    \caption{ Two kinds of proposals for $n_\ell$ which satisfy $(dn)_p=0$. In
        red is a local update which is an exact form, i.e. the exterior derivative $dz$ of a single-site
        0-form. The blue and green updates consist of non-contractible strips of links, and are
        closed because opposite edges of a plaquette contribute inversely to
        $(dn)_p$. }
    \label{fig:constrained-link-updates}
\end{figure}

As before, we offer two kinds of proposals, depicted in
Figure~\ref{fig:constrained-link-updates}, which together reach all
constraint-satisfying configurations of $n$.

We first an offer update of the closed $1$-form $w$.
We update a torus-wrapping strip of parallel links at once, as shown by the blue and green updates in Figure~\ref{fig:constrained-link-updates}.
We pick a single integer $\Delta w \in [-W, +W]$ with $W\in\ZZ$ and Metropolis test $n_\ell \to n_\ell + \Delta w$ for all the links on the strip.
We can sweep this update across all the strips of the lattice in either orientation.

The second proposal offers a local update to the exact $1$-form $dz$.
We pick $\Delta z \in [-Z, +Z]$ with $Z\in\ZZ$ and a site $s$ and build a 0-form $z$ that vanishes everywhere except at $s$ where it is $-\Delta z$.
We propose $n_\ell \to n_\ell + (dz)_\ell$, which amounts to the simultaneous proposal
\begin{align}
    n_{s,\hat{0}} &\to n_{s,\hat{0}} + \Delta z
    &
    n_{s-\hat{0},\hat{0}} &\to n_{s-\hat{0},\hat{0}} - \Delta z
    \nonumber\\
    n_{s,\hat{1}} &\to n_{s,\hat{1}} + \Delta z
    &
    n_{s-\hat{1},\hat{1}} &\to n_{s-\hat{1},\hat{1}} - \Delta z
\end{align}
where $\hat{0}$ and $\hat{1}$ are unit vectors in the positive space and time directions.

An ergodic algorithm should offer proposals of both kinds, and as before their
relative frequency may be adjusted to control autocorrelation times. The
distributions of $\Delta w$ and $\Delta z$ may be adjusted to optimize
performance, for example by changing $W$ and $Z$.

The constrained update algorithms presented here are simple examples of field
update methods which evade the sign problem. We again emphasize that the problem
of constructing ergodic detailed-balanced algorithms to sample discrete gauge
fields satisfying flatness constraints has long been solved in the literature,
and we have only described the algorithms above to make our paper
self-contained.  The specific algorithms we have presented might have long
autocorrelation times in some parameter regimes, especially given that some of
the proposals touch a number of variables growing with the lattice size $N$ and
may be rejected often.  In practice one may want to construct more efficient
constraint-satisfying field update algorithms.  For example, worm algorithms
\cite{PROKOFEV1998253,PhysRevLett.87.160601} can quickly decorrelate worldline
formulations which have closed-loop constraints; it would be interesting to try
and adapt these powerful tools to our actions.

\section{2d QED with $N_f = 2$} 

Let us now consider 2d QED with two flavors of
massless Dirac fermions $\psi, \hat{\psi}$ with a common charge $Q$. The global
flavor symmetry is
\begin{align}
    G_{N_f=2}
    = 
    \frac{SU(2)_L \times SU(2)_R}{\ZZ_2} 
    \times (\ZZ_2)_G
    \times (\ZZ_Q)_A
\nonumber
\end{align}
where $SU(2)_{L,R}$ act on the left and right-handed components of $\psi,
\hat{\psi}$, the quotient is by the gauge transformation $\psi, \hat{\psi} \to
-\psi, -\hat{\psi}$, $(\ZZ_2)_G$ is $G$-parity~\cite{Lee:1956sw}, and the
discrete axial symmetry $(\ZZ_Q)_A$ is the same as before. There is also a
$(\ZZ_Q)_e$ 1-form symmetry. This model is believed to be equivalent to a
self-dual $c=1$ compact boson CFT plus a decoupled massive Schwinger
boson~\cite{Coleman:1975pw,Coleman:1976uz,Gepner:1984au,Affleck:1985wa}. Mass
terms and other perturbations can make this model strongly coupled, and so this
field theory has been a popular testing ground for analytic and numeric
approaches to confining gauge
theories~\cite{Harada:1993va,Shifman:1994ce,Hetrick:1995wq,
Delphenich:1997ex,Narayanan:2012qf,Lohmayer:2013eka,Tanizaki:2016xcu,
Anber:2018jdf,Anber:2018xek,Armoni:2018bga,Misumi:2019dwq,
Georgi:2020jik,Georgi:2022sdu,hirtler2022massless,Gattringer_2015,
Albergo:2022qfi,hirtler2022massless,Delmastro:2021otj,
Delmastro:2022prj,Dempsey:2023gib,Dempsey:2022nys,
Okuda:2022hsq,Itou:2023img,Castellanos:2023rdw,Funcke:2023lli,Ohata:2023sqc}.

Abelian bosonization maps $\psi, \hat{\psi}$ to a pair of $2\pi$-periodic
compact bosons $\varphi, \hat{\varphi}$, so we can discretize it in a parallel
way to the  one-flavor case \eqref{eq:lattice_action}:
\begin{align}
    S_{N_{f}=2}
    &=
    \frac{\beta}{2}  \left[(da)_{p} - 2\pi r_p\right]^2
    \label{eq:two_flavor_QED} \\
    &+
   \frac{\kappa}{2} \left(
        \left[(d\varphi)_{\star\ell} - 2\pi n_{\star\ell}\right]^2 
    +   \left[(d\hat{\varphi})_{\star\ell} - 2\pi \hat{n}_{\star\ell}\right]^2
    \right)
    \nonumber  \\
    &+ \frac{i Q}{2\pi} \left( \varphi_{\star p} +  
    \hat{\varphi}_{\star p} \right) \left[(da)_p - 2\pi r_p \right]
    \nonumber\\
    &
    -   i Q(n_{\star \ell} + \hat{n}_{\star \ell}) a_{\ell} 
    +   i n_{\star \ell} (d \chi)_{\ell}
    +   i \hat{n}_{\star \ell} (d\hat{\chi})_{\ell} \,.
    \nonumber 
\end{align}
where $a,\varphi, \hat{\varphi}, \chi, \hat{\chi} \in \RR$, $n, \hat{n},r \in
\ZZ$, and gauge transformations act as in the one-flavor case
\eqref{eq:one_flavor_gauge} plus the analogous shifts of $\hat{\varphi}$,
$\hat{n}$, and $\hat{\chi}$ with $\hat{k}, \hat{h} \in \ZZ$. 

As before, the lattice parameter $\beta=1/(2e^2a^2)$ and the continuum limit
requires the same scaling as in $N_f = 1$ QED. However, we will see below that
now $\kappa = 1/4\pi$ \emph{is} associated with an enhanced symmetry of the
action \eqref{eq:two_flavor_QED}, and thus we can set $\kappa = 1/4\pi$ on the
lattice and be sure that the lattice theory will flow precisely to $N_f=2$
charge $Q$ massless QED in the continuum limit without any Thirring terms.

The abelian subgroup of $G_{N_f = 2}$ is manifestly respected by the two-flavor
action \eqref{eq:two_flavor_QED}, and we will argue below that the theory flows
to a continuum limit where all of $G_{N_f=2}$ is preserved.   Following a
similar dualization procedure to the $N_f=1$ case (see Appendix
\ref{sec:2flavor_appendix}) we reach
\begin{align}
    S_{N_{f}=2,\textrm{ dual}}
    &
    =   \frac{1}{4\kappa(2\pi)^2} \left[(d\sigma)_{\ell} - 2\pi u_{\ell} \right]^2
    +   i \phi_{\star p} (du)_{p}
    \nonumber\\
    &
    +   \frac{\kappa}{4}\left[(d\eta)_{\tell} - \frac{2\pi}{Q}(dt)_{\tell}\right]^2
    -   \frac{2\pi i}{Q} \hat{u}_{\ell} (dt)_{\star \ell} 
    \nonumber \\
    &
    +   \frac{1}{2\beta}\left(\frac{Q}{2\pi}\right)^2 
    \left( \eta_{\ts} - \frac{2\pi}{Q}t_{\ts}\right)^2
    \label{eq:two_flavor_dual}
\end{align}
where  the fields $u, \hat{u}, t \in \ZZ$ emerge during the dualization process,
and
\begin{align} \label{eq:2flavor_dual_variables} 
&\sigma =\chi -  \hat{\chi}\,,\ \eta = \varphi + \hat{\varphi}\,,\ 
\phi = \frac{\hat{\varphi}}{2}  - \frac{\varphi}{2} - \frac{\pi}{Q} t \nonumber
\end{align}
are real and invariant under $U(1)$ gauge transformations. The remaining gauge redundancies are
\begin{subequations}
\begin{align}
\sigma_s &\to \sigma_s + 2\pi h_s\,, \ u_\ell \to u_\ell + (dh)_\ell\,, \\
\eta_{\tilde s} &\to \eta_{\tilde s} + 2\pi k_{\tilde s}\,,
 \ t_{\tilde s} \to t_{\tilde s}  + Q k_{\tilde s}\,, \\
\phi_{\tilde s} &\to \phi_{\tilde s} + 2\pi \hat h_{\tilde s}\,, 
\ \hat u_{\ell} \to \hat u_{\ell} + (dw)_\ell + Q g_\ell\,,
\end{align}
\end{subequations}
with all gauge parameters taking values in $\ZZ$. Finally, the terms with
factors of $i$ simply impose constraints $(du)_p = 0$ and $(dt)_{\tilde\ell} =
0$ mod $Q$, and solving these constraints when generating field configurations
in a Monte Carlo calculation avoids the sign problem, as discussed in the
preceding section.

The dual formulation \eqref{eq:two_flavor_dual} shows that in the massless
limit, our lattice theory decomposes into two decoupled sectors. The top line of
\eqref{eq:two_flavor_dual} is simply the modified Villain discretization of the
compact boson $\sigma$. But when we set  $\kappa = 1/(4\pi)$, the effective
radius of $\sigma$ makes the theory self-dual under Poisson resummation on
$u_{\ell}$, which implements T-duality on the lattice~\cite{Gorantla:2021svj}.
This implies the existence of a topological line operator which is absent for
generic $\kappa$~\cite{Kapustin:2009av,Choi:2021kmx}, so that the $\kappa =
1/(4\pi)$ point is protected by an enhanced symmetry against quantum
corrections! The continuum limit is thus guaranteed to be the self-dual $c=1$
compact boson CFT with non-abelian $[SU(2) \times SU(2)]/\ZZ_2$ global symmetry.
The decoupled `Schwinger boson' QFT in the lower lines of the dual action
\eqref{eq:two_flavor_dual} matches the remaining symmetries: 
\begin{align}
(\ZZ_2)_G
    &:  \eta_{\ts} \to -\eta_{\ts}, \  \phi_{\ts} \to 
    \phi_{\ts} + \frac{2\pi}{Q} t, \nonumber\\
    & \quad \! \hat{u}_{\ell} \to  - \hat{u}_{\ell} + u_{\ell}, \ t_{\ts}\to -t_{\ts} \\
(\ZZ_Q)_A
    &: \eta_{\tilde s} \to \eta_{\tilde s} + \frac{2\pi q}{Q}, 
    \ t_{\tilde s} \to t_{\tilde s} + q_{\tilde{s}}, q \in \ZZ
    \nonumber\\
(\ZZ_Q)_e
    &:\hat u_\ell \to \hat u_\ell + \frac{2\pi}{Q}v_\ell,
     \ v_\ell \in \ZZ, (dv)_p  =0\,. \nonumber
\end{align}

\section{Chiral gauge theory}  
\label{sec:3450_model}
We now turn to a popular
example~\cite{Halliday:1985tg,Eichten:1985ft,Narayanan:1996kz,
Kikukawa:1997dv,Kikukawa:1997md,Bhattacharya:2006dc,Giedt:2007qg,
Chen:2012di,Poppitz:2007tu,Poppitz:2009gt,Poppitz:2010at,Chen:2012di,
Wang:2013yta,Wang:2018ugf,Tong:2021phe,Zeng:2022grc,Wang:2022ucy,Wang:2022fzc,Lu:2022qtc,
Seifnashri:2023dpa} of a 2d abelian chiral gauge theory, namely the ``$3450$''
model, which has two left-handed Weyl fermions $\psi_{L},\hat{\psi}_{L}$ coupled
to a $U(1)$ gauge field with charges $3,4$  as well as two right-handed Weyl
fermions $\psi_{R}, \hat{\psi}_{R}$ with charges $5,0$.\footnote{In particular
\cite{Seifnashri:2023dpa} mentioned that a Villain
Hamiltonian~\cite{Fazza:2022fss,Cheng:2022sgb} formulation of the 3450 model
should exist. } This QFT satisfies the gauge anomaly cancellation condition $
(Q_L)^2 + (\hat{Q}_L)^2= (Q_R)^2 + (\hat{Q}_R)^2$ as well as the gravitational
't Hooft anomaly cancellation condition on the left and right central charges
$c_L = c_R$. After repackaging  the matter into two Dirac fermions $\psi =
(\psi_{R}, \psi_{L})\transpose,\hat{\psi} = (\hat{\psi}_{R},
\hat{\psi}_{L})\transpose$,  the gauge field couples to the vector and axial
currents of $\psi, \hat{\psi}$ with charges 
$
    Q_{V} = 8, Q_{A} = -2, \hat{Q}_{V}  = 4, \hat{Q}_{A} = 4\,,
$
and  the gauge anomaly cancellation condition becomes $Q_V Q_A + \hat{Q}_V \hat{Q}_A = 0$.

We will study the variant of the $3450$ model with a gauged $(-1)^F$ symmetry to avoid dealing with the Arf invariant~\cite{Thorngren:2018bhj,Karch:2019lnn}.
Our discretization takes the form
\begin{align}
  S_{3450} &= \frac{\beta}{2}  \left[(da)_{p} - 2\pi r_p\right]^2
  \label{eq:3450} \\
  &+  \frac{\kappa}{2} \bigg(
     \left[(d\varphi)_{\tell} - Q_A a_{f(\tell)}-  2\pi n_{\tell}\right]^2 \nonumber \\
     &\phantom{=\frac{\kappa}{2}\bigg(}+\left[(d\hat{\varphi})_{\tell} - \hat{Q}_A a_{f(\tell)}- 2\pi \hat{n}_{\tell}\right]^2 \bigg)
  \nonumber \\
&+ \frac{i }{2\pi} \left(Q_V \varphi_{\star p} +  \hat{Q}_V \hat{\varphi}_{\star p} \right) 
\left[(da)_p 
- 2\pi r_p \right]    \nonumber\\
  &
  - i (Q_V n_{\star \ell} + \hat{Q}_V\hat{n}_{\star \ell}) a_{\ell} 
  + i n_{\star \ell} (d \chi)_{\ell}
  + i \hat{n}_{\star \ell} (d\hat{\chi})_{\ell} \nonumber  \\
  &- i  r_{f(\star s)}(Q_A \chi_s +\hat Q_A \hat \chi_s)    \,.
  \nonumber 
\end{align}
where  $f: s \to s + \tfrac{1}{2}(\hat x + \hat y)$ shifts cells from the lattice to the dual lattice, and the gauge redundancies are 
\begin{align}
a_{\ell} &\to a_{\ell} + (d\lambda)_{\ell} + 2\pi m_{\ell}\,,
\;
r_p \to r_p + (dm)_p 
\label{eq:3450_gauge}\\
  \varphi_{\tilde{s}}  &\to \varphi_{\tilde{s}} + Q_A \lambda_{f(\tilde s)}+ 2\pi k_{\tilde{s}}, 
  \hat{\varphi}_{\tilde{s}} \to \hat{\varphi}_{\tilde{s}}  + \hat{Q}_A \lambda_{f(\tilde s)} +2\pi \hat{k}_{\tilde{s}} \nonumber\\
 n_{\tilde{\ell}} &\to n_{\tilde{\ell}} + (dk)_{\tilde{\ell}} - Q_A m_{f(\tell)}\,, 
 \hat{n}_{\tilde{\ell}} \to \hat{n}_{\tilde{\ell}} + (d\hat{k})_{\tilde{\ell}} - \hat{Q}_A m_{f(\tell)} \nonumber\\
 \chi_s &\to \chi_s + Q_V \lambda_s + 2\pi h_s \,, \hat{\chi}_s \to \hat{\chi}_s + \hat{Q}_V \lambda_s + 2\pi \hat{h}_s \,.\nonumber
\end{align}
Modulo  $2\pi i$ and total derivative terms, the gauge variation of $S_{3450}$ is
\begin{align} \label{eq:chiral_variation}
 & \Delta S_{3450} = i (Q_VQ_A+\hat Q_V\hat Q_A)\Bigg[m_\ell a_{f^{-1}(\star\ell)} \\
 &+ \lambda_s\Big(\frac{1}{2\pi}(da)_{f^{-1}(\star s)}
   - r_{f^{-1}(\star s)} - r_{f(\star s)} - (dm)_{f(\star s)} \Big)  \Bigg]    \nonumber 
\end{align}
which vanishes precisely when the charges satisfy the anomaly cancellation condition.

The function $f$ was introduced to allow the $U(1)$ gauge field to couple to fields that live on the `primary' lattice as well as on the dual lattice, which is necessary in the present context when trying to couple the gauge field to both vector and axial currents. But the presence of the function $f$ in \eqref{eq:3450} appears to break
$\mathbb{Z}_4$ lattice rotation symmetry.  Also, $\textrm{Im} S_{3450} \neq 0$,
leading to an apparent sign problem. However, following the same method as  in
$N_f=1,2$ vector-like QED, one can derive (see Appendix \ref{sec:3450_appendix})
a dual representation which both avoids the sign problem and shows that
$\mathbb{Z}_4$ lattice rotation symmetry is actually preserved, since it is
manifest in the dual variables.  This dual representation can be written as
\begin{multline}
    S = \frac{\kappa}{2}\frac{1}{5} \left( (d\phi)_{\star\ell} -2\pi v_{\star\ell} \right)^2 \\
     + \frac{1}{2\kappa}\frac{1}{20(2\pi)^2}\left( 2(d\hat{\psi})_{\tilde{\ell}} - 2\pi ((dy)_{\tilde{\ell}}- 4v_{\tilde{\ell}}) \right)^2 \\
     + \frac{1}{2\beta}\frac{1}{(2\pi)^2} (4\phi_{\ts} + 2\hat{\psi}_{\ts} - 2\pi y_{\ts})^2 \\
     + i \sigma_{\star\tilde p} (dv)_{\tilde p}  - i\pi \hat n_{\star\tilde \ell} (dy)_{\tilde\ell}   - 2\pi i \frac{4}{5}v_{\star\ell}v_{f(\ell)}\,.
    \end{multline}
The fields $\phi$ and $\hat\psi$ are $U(1)$-gauge invariant combinations of the
fields in \eqref{eq:3450}:
\begin{align}
    \phi_{\ts} = 2\varphi_{\ts}+
\hat{\varphi}_{\ts}\,,\ 
\hat\psi_{\ts} = 2\hat{\chi}_{f(s)} - \chi_{f(s)}\,.
\end{align}  
When
$\hat\psi_{\ts} \to \psi_{\ts} + 2\pi q_{\ts}$ and $\phi_{\tilde s} \to \phi_{\tilde
s} + 2\pi b_{\tilde s}$, the discrete fields shift as $v_{\tilde\ell} \to
v_{\tilde \ell} + (db)_{\tilde\ell}$ and $y_{\tilde s} \to y_{\tilde s} + 4
b_{\tilde s} + 2 q_{\tilde s}$.

We can simplify the expression above further by introducing the variable
\begin{align}
    \rho_{\ts} \equiv  \frac{\pi}{2} y_{\ts} - \frac{1}{2}\hat{\psi}_{\ts} 
    =  \frac{\pi}{2} y_{\ts} + \frac{1}{2}\chi_{f(s)} - \hat{\chi}_{f(s)}
\end{align}
which shifts as $\rho_{\ts} \to
\rho_{\ts} + 2\pi b_{\ts}$ under the discrete gauge transformations described in
the preceding paragraph.  We can then rewrite the dualized action of the
discretized 3450 chiral gauge theory as
\begin{align}
    S &= \frac{\kappa}{10}\left[ (d\phi)_{\tilde\ell} -2\pi v_{\tilde\ell} \right]^2 
    + \frac{1}{10\pi^2 \kappa}\left[
        (d\rho)_{\tilde{\ell}} - 2\pi v_{\tilde{\ell}} \right]^2
        \nonumber \\
        &+ \frac{2}{\pi^2\beta} \left(\phi_{\ts} - \rho_{\ts}\right)^2 +
        i \sigma_{\star\tilde p} (dv)_{\tilde p}  
         - i\pi \hat n_{\star\tilde \ell} (dy)_{\tilde\ell}
        \label{eq:dual_3450}
        \\
        &  - 2\pi i \frac{4}{5}v_{\star\ell}v_{f(\ell)} \,. \nonumber
\end{align}
This action has several interesting physical consequences.  First, the $i \pi
\hat{n} \, dy$ term is a decoupled topological $\mathbb{Z}_2$ gauge theory.  This
TQFT appears because the charges of the Dirac fermions in the 3450 chiral gauge
theory are even with a minimal charge of $2$.  Therefore the model has a
$\mathbb{Z}_2$ $1$-form symmetry, and this symmetry is spontaneously broken.
Second, we see that one linear combination of $\phi$ and $\rho$ acquires a
Schwinger-type mass.  These two features are analogous to the behavior of the
charge-$Q$ $N_f=1$ Schwinger model.  Third, another linear combination of $\phi$
and $\rho$ remains exactly massless, thanks to the presence of the $\sigma\, dv$
term, which ensures that there are no dynamical $v$ vortices, implying that
there cannot be any BKT transition~\cite{Berezinsky:1970fr,Kosterlitz:1973xp} as
a function of $\kappa$. This gapless mode matches the $U(1) \times U(1)$ 't
Hooft anomaly of the 3450 model, which is analogous to what we saw above in the
$N_f=2$ vector-like QED.

The last interesting physical consequence of Eq.~\eqref{eq:dual_3450} we want to
highlight is that the model has an extra $\mathbb{Z}_2$ $0$-form symmetry that
acts by exchanging $\phi$ and $\rho$ if we
set $\kappa = 1/\pi$! Dialing $\kappa$ maps to dialing the coefficients of the
Thirring interaction terms $(\bar{\psi} \gamma^{\mu}\psi)^2$ and $(\bar{\hat\psi}
\gamma^{\mu}\hat\psi)^2$ in the original fermionic theory.  The extra $\mathbb{Z}_2$
symmetry is therefore not present at weak coupling, helping to explain why it is not
obvious in the original fermionic description of the model. It is even quite
opaque after bosonization, and only becomes obvious after finding a particularly
simple duality frame.  Another reason this symmetry is not obvious from the start
is that it involves exchanging $\phi$, a six-fermion operator, with $\rho$,
which is an exotic defect operator from the point of view of the original
fermionic description.

Before closing this section, we can return to our original motivations for
looking for a dual representation of Eq.~\eqref{eq:3450}: the sign problem and
the lack of a manifest $\mathbb{Z}_4$ rotation symmetry.  We have already seen
that the latter issue is automatically taken care of by passing to the
representation in Eq.~\eqref{eq:dual_3450}, so all that remains is understanding
why in the end there is no sign problem.  The
first two terms in the second line of Eq.~\eqref{eq:dual_3450} yield constraints
$(dv)_{\tilde p} = 0$ and $(dy)_{\tilde\ell} = 0$ mod $2$.  We have already seen
that such constraints are easy to enforce when generating field configurations,
and consequently these terms are harmless.  The remaining term in the bottom of
Eq.~\eqref{eq:dual_3450} looks alarming at first glance, but when $dv = 0$, it
can be shown to be a total derivative (see
Appendix~\ref{sec:3450_appendix}) and can be dropped, giving a sign-problem-free
formulation with a manifest $\mathbb{Z}_4$ rotation symmetry.

\section{Outlook}
We leveraged recent advances in the understanding of anomalies on the
lattice~\cite{Sulejmanpasic:2019ytl,Gorantla:2021svj,Fazza:2022fss,
Cheng:2022sgb,Seifnashri:2023dpa} to construct symmetry-preserving
discretizations of $d=2$ abelian gauge theories with massless fermions with
vector-like and chiral couplings. These discretizations evade the
Nielsen-Ninomiya theorem essentially because they directly discretize the
fermion determinant $\det \slashed{D}$ rather than the fermion matrix
$\slashed{D}$ itself, and $\det \slashed{D}$ is rewritten in terms of a path
integral over bosonic fields with local interactions.  We emphasize that in our
construction, the vector-like and chiral symmetries act just as locally at
finite lattice spacing as they do in the continuum, and all of the ABJ and 't
Hooft anomalies are reproduced on the lattice. Finally, while
the lattice actions we construct are complex, we have shown that the resulting
sign problems can be avoided by judicious choices of dual variables.

Our results open many directions for future work.  Numerical lattice
calculations using this formalism can be used to explore strongly-coupled
regions in parameter space. It would be interesting to see if our approach can
be generalized to $d>2$, for example by taking advantage of advances in the
understanding of continuum bosonization in
$d=3$~\cite{Aharony:2015mjs,Seiberg:2016gmd,Karch:2016sxi,Karch:2016aux,Wang:2017txt,Hsin:2016blu,Karch:2018mer}
and the development of symmetry-preserving discretizations of Chern-Simons
terms~\cite{Jacobson:2023cmr}.  It would also be nice to see if  generalizations
of our construction can preserve non-Abelian chiral symmetries at finite lattice
spacing~\cite{Luscher:1999un,Wen:2013ppa,BenTov:2014eea,Ayyar:2014eua,Ayyar:2015lrd,Ayyar:2016lxq,DeMarco:2017gcb,Razamat:2020kyf,Catterall:2022jky}.
Finally, to get  inspiration towards constructing more direct
symmetry-preserving fermion discretizations, one can compute the discretized
Dirac operators corresponding to our representations of the fermion determinant
$\det \slashed{D}$.

{\bf Acknowledgements.} We thank Shi Chen, Maria Neuzil, Erich Poppitz, Srimoyee Sen and Misha Shifman for
helpful discussions, are grateful to Srimoyee Sen and Juven Wang for comments on a draft, and thank the audience at the Simons Foundation ``Confinement and QCD Strings" 2023 Annual Meeting for helpful remarks during a presentation of our results. The
work of EB and AC was performed in part at the Aspen Center for Physics, which
is supported by National Science Foundation grant PHY-2210452. AC was also
supported by a Simons Foundation Grant No. 994302 as part of the Simons
Collaboration on Confinement and the QCD String. TJ is supported by a Julian
Schwinger Fellowship from the Mani L. Bhaumik Institute for Theoretical Physics
at UCLA, and thanks NCSU for hospitality during the completion of this work.

\bibliography{non_inv}

\newpage

\appendix

\section{Hamiltonian Formulation}
\label{sec:HamiltonianAppendix}

To construct the Hamiltonian for the 1-flavor bosonized Schwinger model we follow the discussion of Ref.~\cite{Fazza:2022fss} (see Section 3.5 therein for a parallel discussion).
We go to Lorentzian signature, take time to be continuous, drop the time-like integer-valued fields in the lattice action \eqref{eq:lattice_action}, and assume that space is discretized on a lattice with periodic boundary conditions.
The Lagrangian density becomes
\begin{align}
    \mathcal{L}
    =   \frac{\kappa}{2} \dot{\varphi}_{\tilde{s}}^2 
    -   \frac{\kappa}{2}[(d\varphi)_{\tilde{\ell}} - 2\pi n_{\tilde{\ell}}]^2
    +   \frac{\beta}{2} \dot{a}^2_{\ell}
    -   \frac{Q}{2\pi} \varphi_{\star \ell} \dot{a}_{\ell}
    +   \chi_{s} \dot{n}_{\star s}
\end{align}
Note that on the 1d lattice $\star \ell = \tilde{s}, \star s = \tilde{\ell}$.
The canonical momenta (which live on the duals of the cells of their respective fields) can thus be written as
\begin{align}
    (\Pi_{\varphi})_{\ell} &= \kappa \dot{\varphi}_{\star \ell}\,, \;
    (\Pi_a)_{\tilde{s}} = \beta \dot{a}_{\ell} - \frac{Q}{2\pi} \\
    (\Pi_n)_{s} &= \chi_s\,, \;
    (\Pi_{\chi})_{\tilde{\ell}} = 0
\end{align}
The lower two lines are second-class constraints.
They can be taken into account using Dirac brackets, which lead us to the quantum Hamiltonian
\begin{align}
    \Hop
    &=  \frac{1}{2\kappa}(\Piphi)_{\ell}^2
    +   \frac{1}{2\beta} \left[(\Pia)_{\tilde{s}}+ \frac{Q}{2\pi}\phiop_{\tilde{s}}\right]^2
    \nonumber\\
    &+  \frac{\kappa}{2}[(d\phiop)_{\tilde{\ell}} - 2\pi \nop_{\tilde{\ell}}]^2
    \label{eq:hamiltonian}
\end{align}
where $\phiop, \Piphi, \aop, \Pia, \nop, \chiop$ are operators with the commutation relations
\begin{align}
    [\phiop_{\tilde{s}},(\Piphi)_{\ell}] &= i \delta_{\tilde{s}, \star\ell},
    \nonumber\\
    [\aop_{\ell},(\Pia)_{\tilde{s}}] &= i \delta_{\ell, \star{\tilde{s}} },
    \label{eq:commutation relations}
    \\\nonumber
    [\nop_{\tilde{\ell}},\chiop_{s}] &= i \delta_{\star \tilde{\ell}, s},
\end{align}

{\bf Gauge operators.}
In the Hamiltonian formalism the gauge redundacies must be imposed as constraints.
We have four such redundancies:  compactness of $\chiop$, compactness of $\phiop$, small gauge transformations of $\aop$ and $\chiop$, and large gauge transformations that ensure that the gauge group is $U(1)$ and not $\RR$.
These transformations will be associated with four operators $\Gchi, \Gphi, \Gsmall, \Glarge$ which have to act like identity operators on all physical states.

The $2\pi$ shifts of $\chi$ are generated by
\begin{align}
    \Gchi
    =
    [\{s_{\tilde{\ell}}\}]
    =
    e^{2\pi i \sum_{\tilde{\ell}} s_{\tilde{\ell}} \nop_{\tilde{\ell}}} \,.
\end{align} 
where $s_{\tilde{\ell}} \in \ZZ$.
Therefore $\nop_{\tilde{\ell}}$ must have an integer spectrum.

The compactness  condition for $\phiop$ is associated with the transformation 
$\phiop_{\tilde{s}} \to \phiop_{\tilde{s}} + 2\pi k_{\tilde{s}},
\nop_{\tilde{\ell}} \to \nop_{\tilde{\ell}} + (dk)_{\tilde{\ell}},
(\Pia)_{\tilde{s}}  \to (\Pia)_{\tilde{s}} - \frac{Q}{2\pi} k_{\tilde{s}}$ with $k_{\tilde{s}} \in \ZZ$, which is generated by
\begin{align}
    \Gphi [\{k\}]
    =
    \exp\left[
        i \sum_{\ell}
        2\pi k_{\star \ell} 
        \left(
            (\Piphi)_{\ell}
            - \frac{Q}{2\pi} \aop_{\ell}
            - \frac{1}{2\pi} (d\chiop)_{\ell}
        \right)
    \right]
\end{align}
Therefore the operator
\begin{align}
    \qop_{\ell} = (\Piphi)_{\ell} + \frac{Q}{2\pi} \aop_{\ell} - \frac{1}{2\pi} (d\chiop)_{\ell}
\end{align}
must have an integer spectrum.
Its commutation relations are
\begin{align}
    [\qop_{\ell}, \phiop_{\tilde{s}}] &= i \delta_{\tilde{s}, \star\ell},
    \nonumber\\
    [\qop_{\ell},(\Pia)_{\tilde{s}}] &= \frac{i Q}{2\pi}\delta_{\ell, \star{\tilde{s}} },
    \\\nonumber
    [\qop_{\ell},\nop_{\tilde{\ell}}] &= \frac{i}{2\pi} \left( \delta_{f(\ell), \tilde{\ell}} - \delta_{f^{-1}(\ell),\tilde{\ell}} \right)
\end{align}
where $f(\ell)$ is a positive translation by half a lattice unit, which takes
the lattice to the dual lattice.

Continuous (`small') gauge transformations $\aop \to \aop + d\lambda, \chiop \to \chiop + Q \lambda$ are generated by
\begin{align}
    \Gsmall(\{\lambda\})
    =
    \exp \left[
        i \sum_{\ell} (d\lambda)_{\ell} (\Pia)_{\ell}
    -   i Q \sum_s \lambda_s \nop_{\star s}
    \right]
\end{align}
where ${\lambda_s} \in \RR$.
This yields the Gauss law constraint
\begin{align}
    (d\Pia)_{\tilde{\ell}} + Q \nop_{\tilde{\ell}} = 0.
\end{align}

Finally, the large gauge transformations $(\Pia)_{\ell} \to (\Pia)_{\ell} + 2\pi m_{\ell}$ are generated by
\begin{align}
  \Glarge(\{m\})
  =
  e^{i \sum_{\ell} 2\pi m_{\ell} (\Pia)_{\ell}} \,, 
  \;
  m_{\ell} \in \ZZ
\end{align}
which implies that $\Pia$ must have an integer spectrum.

{\bf Symmetry operators.}
The two internal global symmetries of our Hamiltonian are associated with the following operators.
The $\ZZ_Q$ chiral symmetry is generated by the line operator
\begin{align}
    \Uchiral_k(L)
    &=
    \exp \left[
        \frac{2\pi i k}{Q} \sum_{\ell \in L} 
        \left((\Piphi)_{\ell} + \frac{Q}{2\pi} (\Pia)_{\ell}\right)
    \right]
    \\
    &=
    e^{\frac{2\pi i k}{Q} \sum_{\ell} \qop_{\ell}} \nonumber
\end{align}
where $L$ is all of space (that is, a time slice).
This means that $\qop_{\ell}$ is a charge density operator (which manages to exist in this case despite the fact that chiral symmetry is discrete) and $\Qop = \sum_{\ell} \qop_{\ell}$ is the total charge operator.
The equation of motion of $\qop$ is $\dot{\qop} = i[\Hop,\qop] = 0$, so $\Uchiral_k(L)$ is conserved, and the coefficient in front of $\Qop$ is quantized thanks to the requirement that $\Uchiral_k(L)$ must commute with $\Glarge$.

The $\ZZ_Q$ $1$-form symmetry is generated by the local operator
\begin{align}
    \Voneform_{w}(\tilde{s})
    =
    e^{\frac{2\pi i }{Q} (\Pia)_{\tilde{s}}}.
\end{align}
The coefficient in the exponent must be quantized so that $[\Voneform_w, \Gphi] = 0$, and it is topological thanks to the Gauss law.

The 't Hooft anomaly is encoded in the fact that these symmetry operators do not commute
\begin{align}
    \Uchiral_k(L) \Voneform_w(\tilde{s})
    =
    e^{\frac{2\pi i k w}{Q}}
    \Voneform_w(\tilde{s})\Uchiral_k(L)
\end{align}
Therefore \eqref{eq:hamiltonian} provides a Hamiltonian discretization of the charge $Q$ Schwinger model which encodes all of its continuum internal symmetries and anomalies.

\section{Dualizing $N_f=2$ QED}
\label{sec:2flavor_appendix}

We start with the action \eqref{eq:two_flavor_QED}, linearize the gauge and scalar kinetic terms using auxiliary fields, and sum over $n, \hat n, r$ to find 
\begin{multline}
\frac{1}{2\kappa}\left(\frac{N}{2\pi}\right)^2\Bigg[ \left(a_\ell -\frac{1}{N}(d\chi)_\ell - \frac{2\pi}{N}y_\ell \right)^2 \\
+ \left(a_\ell -\frac{1}{N}(d\hat\chi)_\ell - \frac{2\pi}{N}\hat u_\ell \right)^2 \Bigg]\\
 + \frac{1}{2\beta}\left(\frac{N}{2\pi}\right)^2 \left( \varphi_{\tilde s} + \hat \varphi_{\tilde s} - \frac{2\pi}{N}t_{\tilde s} \right)^2 + i a_\ell (dt)_{\star p}\\
-\frac{i}{2\pi}(N a_\ell + 2\pi y_\ell)(d\varphi)_{\star\ell}-\frac{i}{2\pi}(N a_\ell + 2\pi \hat u_\ell)(d\hat\varphi)_{\star\ell}\,,
\end{multline}
where $y, \hat u, t \in \ZZ$. Doing the Gaussian integral over $a$ gives
\begin{multline}
\frac{1}{4\kappa}\frac{1}{(2\pi)^2} \left( (d\chi)_\ell - (d\hat\chi)_\ell - 2\pi (y_\ell - \hat u_\ell) \right)^2 \\
+ \frac{\kappa}{4}\left((d\varphi)_{\tilde\ell} + (d\hat\varphi)_{\tilde\ell} - \frac{2\pi}{N}(dt)_{\tilde\ell} \right)^2 \\
+ \frac{1}{2\beta}\left(\frac{N}{2\pi}\right)^2 \left(\varphi_{\tilde s} +\hat\varphi_{\tilde s} - \frac{2\pi}{N}t_{\tilde s} \right)^2 \\
- \frac{i}{2}(y_\ell - \hat u_\ell )( (d\varphi)_{\star\ell} - (d\hat\varphi)_{\star\ell}) -\frac{i}{2}\frac{2\pi}{N}(y_\ell + \hat u_\ell) (dt)_{\star\ell}
\end{multline}
after dropping total derivatives and multiples of $2\pi i$. Now let us define $\sigma =  \chi - \hat \chi$, $\eta = \varphi + \hat\varphi$, $\phi = \frac{\hat\varphi}{2} - \frac{\varphi}{2} - \frac{\pi t}{N}$, $u = y - \hat u$.  
\begin{multline}
\frac{1}{4\kappa}\frac{1}{(2\pi)^2} \left( (d\sigma)_\ell - 2\pi u_\ell \right)^2 \\
+ \frac{\kappa}{4}\left((d\eta)_{\tilde\ell} - \frac{2\pi}{N}(dt)_{\tilde\ell} \right)^2 
+ \frac{1}{2\beta}\left(\frac{N}{2\pi}\right)^2 \left(\eta_{\tilde s} - \frac{2\pi}{N}t_{\tilde s} \right)^2 \\
+i u_\ell  (d\phi)_{\star\ell} -\frac{2\pi i}{N}\hat u_\ell (dt)_{\star\ell}
\end{multline}
which is the result \eqref{eq:two_flavor_dual}. The first line is the modified Villain formulation of a compact scalar. If we had started with the free-fermion radius $\kappa = \frac{1}{4\pi}$, we end up with an effective radius $\tilde\kappa =\frac{1}{2\kappa} \frac{1}{(2\pi)^2} = \frac{1}{2\pi}$, which is the self-dual radius.  

For completeness, we note the dual description of the fermion mass terms: 
\begin{equation}
\cos(\varphi) \to \cos\left(\frac{\eta}{2}-\phi - \frac{\pi}{N}t \right)
\end{equation}
and
\begin{equation}
\cos(\hat\varphi) \to \cos\left(\frac{\eta}{2}+\phi + \frac{\pi}{N}t \right)\,.
\end{equation}
As a result, a flavor-symmetric mass deformation becomes
\begin{equation}
 \cos(\varphi) + \cos(\tilde\varphi)  \to 2\cos\left(\frac{\eta}{2}\right)\cos\left(\phi + \frac{\pi}{N}t\right)\,.
\end{equation}
Note this is respects the $2\pi$ periodicity because when $\eta \to \eta + 2\pi k$ and $t \to t + Nk$, both factors pick up a sign $(-1)^k$ which squares away. If we turn on a theta angle this is modified to $\cos(\frac{\eta}{2} + \frac{\theta}{2})\cos(\phi + \frac{\pi}{N}t)$.

\section{Dualizing the 3450 model}
\label{sec:3450_appendix}

Let us write the action of the 3450 model using (real) auxiliary fields $\zeta,\hat\zeta,\xi$:
\begin{multline}
S = \frac{1}{2\kappa}\zeta_{\star\tell}^2 + i \zeta_{\star\tell}[ (d\varphi)_{\tell} - Q_A\, a_{f(\tell)} - 2\pi n_{\tell}] \\
+\frac{1}{2\kappa}\hat\zeta_{\star\tell}^2 + i \hat\zeta_{\star\tell}[ (d\hat\varphi)_{\tell} - \hat Q_A\, a_{f(\tell)} - 2\pi \hat n_{\tell}]  \\
+\frac{1}{2\beta}\xi_{\star p}^2 + i \left(\xi_{\star p} + \frac{1}{2\pi}(Q_V\varphi_{\star p} +\hat Q_V\hat\varphi_{\star p})\right)[(da)_p -2\pi r_p] \\
-i (Q_V n_{\star\ell} + \hat Q_V \hat n_{\star\ell})a_\ell   + i n_{\star \ell} (d \chi)_{\ell}
  + i \hat{n}_{\star \ell} (d\hat{\chi})_{\ell}  \\
  - i  r_p (Q_A \chi_{f^{-1}(\star p)} +\hat Q_A \hat \chi_{f^{-1}(\star p)} )    \,.
\end{multline}
Summing over $r_p \in \ZZ$ sets 
\begin{multline} \label{eq:xi} 
\xi_{\star p} = -\frac{1}{2\pi}\Big( Q_V \varphi_{\star p} + \hat Q_V\hat \varphi_{\star p} \\
+ Q_A \chi_{f^{-1}(\star p)} + \hat Q_A \hat\chi_{f^{-1}(\star p)} - 2\pi y_{\star p}\Big)\,,
\end{multline}
where $y_{\tilde\ell} \in \ZZ$. Plugging this back into the action and integrating out the gauge field $a_\ell$ allows us to solve for $\zeta$:
\begin{multline}
\zeta_{\star\tell} = \frac{1}{Q_A}\Big[ -\hat Q_A \hat \zeta_{\star\tell} -\frac{1}{2\pi}(Q_A(d\chi)_{\star\tell} + \hat Q_A(d\hat\chi)_{\star\tell})\\
 + (dy)_{f(\star\tell)} - Q_V n_{f(\star\tell)} - \hat Q_V \hat n_{f(\star\tell)} \Big]\,.
\end{multline}
If we plug this back into the action and perform a field redefinition $\hat\zeta_\ell \to \hat\zeta_\ell -\frac{1}{2\pi}(d\hat\chi)_\ell$, we land on 
\begin{multline}
S = \frac{1}{2\beta}\frac{1}{(2\pi)^2}\xi_{\ts}^2 + \frac{1}{2\kappa}\left[\hat \zeta_\ell - \frac{1}{2\pi}(d\hat\chi)_\ell \right]^2\\
+ \frac{1}{2\kappa}\frac{\hat Q_A^2}{ Q_A^2} \left[\hat \zeta_\ell + \frac{Q_A}{\hat Q_A}\frac{(d\chi)_\ell}{2\pi} - \frac{(dy)_{f(\ell)} + u_{f(\ell)}}{\hat Q_A} \right]^2 \\
-i \frac{\hat\zeta_\ell }{Q_A} \left[Q_A (d\hat\varphi)_{\star\ell} - \hat Q_A (d\varphi)_{\star\ell} + 2\pi \frac{\hat Q_A}{Q_V}u_{\star\ell} \right] \\
+ \frac{i}{Q_A} u_{f(\ell)}[(d\varphi)_{\star\ell} - 2\pi n_{\star\ell}]+\frac{2\pi i}{Q_A}(dy)_{f(\ell)}n_{\star\ell} \,,
\end{multline}
with $\xi$ set to its value in \eqref{eq:xi}, and for notational convenience we have defined
\begin{equation}
u_{\tell} \equiv Q_V n_{\tell} + \hat Q_V \hat n_{\tell} = \frac{Q_V}{\hat Q_A}(\hat Q_A n_{\tell} - Q_A\hat n_{\tell})\,,
\end{equation}
where the second equality follows from the anomaly-free condition. Note this is not a $GL(2,\ZZ)$ change of basis, and we have to remember that $u_{\tell} \in Q_V \ZZ + \hat Q_V\ZZ$. The integral over $\hat\zeta$ is Gaussian. To simplify the calculation let us define
\begin{align}
\phi_{\ts} &= Q_V \varphi_{\ts} + \hat Q_V\hat\varphi_{\ts}\,,\; \psi_s = Q_A\chi_s +\hat Q_A \hat \chi_s\,, \\
\eta_s &= \hat Q_A\chi_s - Q_A \hat\chi_s\,.
\end{align} 
Note that $\phi, \psi$ are (0-form) gauge-invariant but $\eta$ is not. One can check that this defines an invertible change of basis assuming the anomaly-cancelation condition. In terms of these new variables, the result of integrating over $\hat\zeta$ is
\begin{multline}
S =  \frac{\kappa}{2}\frac{\hat Q_A^2/Q_V^2}{Q_A^2 + \hat Q_A^2} \left( (d\phi)_{\star\ell} -2\pi  u_{\star\ell}) \right)^2 \\
 + \frac{1}{2\kappa}\frac{1}{(2\pi)^2}\frac{1}{Q_A^2 + \hat Q_A^2}\left( (d\psi)_\ell - 2\pi ((dy)_{f(\ell)}- u_{f(\ell)}) \right)^2 \\
 +\frac{1}{2\beta}\frac{1}{(2\pi)^2} (\phi_{\ts} + \psi_{f^{-1}(\ts)} - 2\pi y_{\ts})^2 \\
  + \frac{i}{Q_A} u_{f(\ell)}[(d\varphi)_{\star\ell} - 2\pi n_{\star\ell}]+\frac{2\pi i}{Q_A}(dy)_{f(\ell)}n_{\star\ell} \\
-\frac{i}{2\pi}\frac{\hat Q_A/Q_V}{Q_A^2+\hat Q_A^2}\left((d\phi)_{\star\ell} -2\pi u_{\star\ell}) \right) \\
\times \left[ (d\eta)_\ell -2\pi \frac{\hat Q_A}{Q_A}\left( (dy)_{f(\ell)}- u_{f(\ell)} \right) \right]
\end{multline}
Let us collect the terms linear in $du$:
\begin{multline}
i (du)_{\tilde p}\Bigg[ \frac{1}{Q_A}\varphi_{f^{-1}(\star\tilde p)} - \frac{\hat Q_A/Q_V}{Q_A^2 + \hat Q_A^2} \eta_{\star\tilde p} \\
-\frac{\hat Q_A^2/(Q_AQ_V)}{Q_A^2+\hat Q_A^2} (\phi_{f^{-1}(\star\tilde p)}- 2\pi y_{f(\star\tilde p)}) \Bigg]
\end{multline} 
Ignoring total derivatives, neither $\varphi$ nor $\eta$ appear anywhere else in the action, so we can freely make another change of variables and define a new field $\sigma_{\star\tilde p}$ which is equal to the quantity in brackets (note that this quantity is gauge invariant!). The role of $\sigma$ is to set $(du)_{\tilde p} = 0$. 

The remaining imaginary terms in the action are
\begin{equation}
\frac{2\pi i}{Q_A}((dy)_{f(\ell)} - u_{f(\ell)}) n_{\star\ell} + 2\pi i\frac{\hat Q_A^2/(Q_AQ_V)}{Q_A^2+\hat Q_A^2} u_{\star\ell} u_{f(\ell)}\,. 
\end{equation}
The last term is trivial when $du = 0$. To see this, fix a plaquette $\tilde p$ on the dual lattice whose lower left-hand corner is at the site $\tilde x$. One can verify the identity\footnote{This is equivalent to the following identity involving higher cup products:
 $2u\cup u = -d(u\cup_1 u) + du\cup_1 u - u \cup_1 du$ where $u$ is an arbitrary 1-form. See~\cite{Jacobson:2023cmr} for explicit formulas for higher cup products on square lattices.}
\begin{multline}
\sum_{\tell=(\tilde x,\mu)} u_{\tell} \, u_{f(\star\tell)} = -\frac{1}{2} ( d(u^2))_{\tilde p} \\
+ \frac{1}{2} (du)_{\tilde p}(u_{\tilde x, 0} + u_{\tilde x + \hat 0,1} + u_{\tilde x, 1} + u_{\tilde x+\hat 1,0})
\end{multline}
where the sum is over the two links emanating from $\tilde x$. The first term is a total derivative which vanishes when summed over the lattice and the second line vanishes when $du =0$. 

Recalling the definition of $u$, we arrive at the dual formulation (ignoring the term discussed above)
\begin{multline}
S = \frac{\kappa}{2}\frac{\hat Q_A^2/Q_V^2}{Q_A^2 + \hat Q_A^2} \left( (d\phi)_{\star\ell} -2\pi  (Q_V n_{\star\ell} + \hat Q_V \hat n_{\star\ell}) \right)^2 \\
 + \frac{1}{2\kappa}\frac{1/(2\pi)^2}{Q_A^2 + \hat Q_A^2}\left( (d\psi)_\ell - 2\pi ((dy)_{f(\ell)}- Q_V n_{f(\ell)} - \hat Q_V \hat n_{f(\ell)}) \right)^2 \\
 + \frac{1}{2\beta}\frac{1}{(2\pi)^2} (\phi_{\ts} + \psi_{f^{-1}(\ts)} - 2\pi y_{\ts})^2 \\
 + i \sigma_{\star\tilde p}(Q_V (dn)_{\tilde p} + \hat Q_V (d\hat n)_{\tilde p}) \\
  + \frac{2\pi i}{Q_A}((dy)_{f(\ell)} - Q_V n_{f(\ell)} - \hat Q_V \hat n_{f(\ell)}) n_{\star\ell}\,.
  \label{eq:remaining_sign}
 \end{multline}
So far we have allowed the charges to be arbitrary, but satisfying the anomaly-cancelation condition. Note that there are still imaginary terms which remain, indicating a sign problem. In future work, it would be interesting to determine the set of 2d chiral gauge theories where one can avoid this sign problem, possibly by finding alternatives to \eqref{eq:remaining_sign} in some cases  

Here we focus on the particular case of the 3450 model, where the sign problem can indeed be removed thanks to the fact that the `3450' charge assignments make the last two terms in \eqref{eq:remaining_sign}  multiples of $2\pi i$.  Dropping these terms we find 
\begin{multline}
S = \frac{\kappa}{2}\frac{1}{80} \left( (d\phi)_{\star\ell} -2\pi  (8 n_{\star\ell} + 4 \hat n_{\star\ell}) \right)^2 \\
 + \frac{1}{2\kappa}\frac{1}{20(2\pi)^2}\left( (d\psi)_\ell - 2\pi ((dy)_{f(\ell)}- 8 n_{f(\ell)} - 4 \hat n_{f(\ell)}) \right)^2 \\
 + \frac{1}{2\beta}\frac{1}{(2\pi)^2} (\phi_{\ts} + \psi_{f^{-1}(\ts)} - 2\pi y_{\ts})^2 \\
 + i \sigma_{\star\tilde p}(8 (dn)_{\tilde p} + 4 (d\hat n)_{\tilde p})  - i\pi (dy)_{f(\ell)}  n_{\star\ell} \\
 - \frac{2\pi i}{20}(8n_{\star\ell} + 4\hat n_{\star\ell})(8n_{f(\ell)} + 4\hat n_{f(\ell)})\,,
 \end{multline}
where for completeness we have reinstated the imaginary term we argued was zero above. Now we can perform a $GL(2,\ZZ)$ change of basis to $v_{\tell} = 2n_{\tell} + \hat n_{\tell}$ and $n_{\tell}$ as the integer degrees of freedom, 
\begin{multline}
S = \frac{\kappa}{2}\frac{1}{80} \left( (d\phi)_{\star\ell} -8\pi v_{\star\ell} \right)^2 \\
 + \frac{1}{2\kappa}\frac{1}{20(2\pi)^2}\left( (d\psi)_\ell - 2\pi ((dy)_{f(\ell)}- 4v_{f(\ell)}) \right)^2 \\
 + \frac{1}{2\beta}\frac{1}{(2\pi)^2} (\phi_{\ts} + \psi_{f^{-1}(\ts)} - 2\pi y_{\ts})^2 \\
 + 4i \sigma_{\star\tilde p} (dv)_{\tilde p}  - i\pi (dy)_{f(\ell)}  n_{\star\ell} - 2\pi i \frac{4}{5}v_{\star\ell}v_{f(\ell)} \,.
 \end{multline}
We now rescale $\phi \to 4\phi$, $\psi \to 2\psi$,  $\sigma \to \frac{1}{4}\sigma$ to reach
\begin{multline}
S = \frac{\kappa}{2}\frac{1}{5} \left( (d\phi)_{\star\ell} -2\pi v_{\star\ell} \right)^2 \\
 + \frac{1}{2\kappa}\frac{1}{20(2\pi)^2}\left( 2(d\psi)_\ell - 2\pi ((dy)_{f(\ell)}- 4v_{f(\ell)}) \right)^2 \\
 + \frac{1}{2\beta}\frac{1}{(2\pi)^2} (4\phi_{\ts} + 2\psi_{f^{-1}(\ts)} - 2\pi y_{\ts})^2 \\
 + i \sigma_{\star\tilde p} (dv)_{\tilde p}  - i\pi (dy)_{f(\ell)}  n_{\star\ell}- 2\pi i \frac{4}{5}v_{\star\ell}v_{f(\ell)}\,.
 \end{multline}
The $\mathbb{Z}_4$ lattice rotation symmetry is not manifest at this level --- for instance the second and third lines involve the shift $f$ which does not commute with rotations. However, the first three lines of the action are invariant if one performs a (counter-clockwise) $\pi/2$ rotation together with a shift $\psi_s \to \psi_{s + \hat 0}$. The last line just encodes the constraints, modulo the last term which we argued is a total derivative. Alternatively, we can make the $\mathbb{Z}_4$ lattice rotation symmetry manifest by simply defining
 $\hat{\psi}_{\tilde{s}} = \psi_{f^{-1}(\ts)}$ and $\hat n_{\tilde\ell} = n_{f^{-1}(\star\tilde\ell)}$, so that
\begin{multline}
S = \frac{\kappa}{2}\frac{1}{5} \left( (d\phi)_{\star\ell} -2\pi v_{\star\ell} \right)^2 \\
 + \frac{1}{2\kappa}\frac{1}{20(2\pi)^2}\left( 2(d\hat{\psi})_{f(\ell)} - 2\pi ((dy)_{f(\ell)}- 4v_{f(\ell)}) \right)^2 \\
 + \frac{1}{2\beta}\frac{1}{(2\pi)^2} (4\phi_{\ts} + 2\hat{\psi}_{\ts} - 2\pi y_{\ts})^2 \\
 + i \sigma_{\star\tilde p} (dv)_{\tilde p}  - i\pi \hat n_{\star\tilde \ell} (dy)_{\tilde\ell}  - 2\pi i \frac{4}{5}v_{\star\ell}v_{f(\ell)}\,.
 \end{multline} 
The path integrals over $\sigma$ and $\hat n$ serve to impose the constraints
$(dv)_{\tilde{p}} = 0, (dy)_{\tilde{\ell}} = 0 \textrm{ mod } 2$, so that
$\sum_{\ell} v_{\star\ell}v_{f(\ell)} = 0$.  Finally, we observe that for any lattice field
$g_{\tilde{\ell}}$, $\sum_{\ell} g_{f(\ell)}^2 = \sum_{\tilde{\ell}}
g_{\tilde{\ell}}^2$, so that we can rewrite $S$ as 
\begin{multline}
S = \frac{\kappa}{2}\frac{1}{5} \left( (d\phi)_{\star\ell} -2\pi v_{\star\ell} \right)^2 \\
 + \frac{1}{2\kappa}\frac{1}{20(2\pi)^2}\left( 2(d\hat{\psi})_{\tilde{\ell}} - 2\pi ((dy)_{\tilde{\ell}}- 4v_{\tilde{\ell}}) \right)^2 \\
 + \frac{1}{2\beta}\frac{1}{(2\pi)^2} (4\phi_{\ts} + 2\hat{\psi}_{\ts} - 2\pi y_{\ts})^2 \\
 + i \sigma_{\star\tilde p} (dv)_{\tilde p}  - i\pi \hat n_{\star\tilde \ell} (dy)_{\tilde\ell}   - 2\pi i \frac{4}{5}v_{\star\ell}v_{f(\ell)}\,.
\end{multline}
This is form of the dual action we will discuss in the main text.

\end{document}